\documentstyle[12pt]{article}
\newcommand{\be}{\begin{equation}}
\newcommand{\ee}{\end{equation}}
\input tcilatex

\begin{document}

\begin{center}
{\Large {\bf A LOCAL HIDDEN VARIABLES MODEL FOR EXPERIMENTS INVOLVING PHOTON
PAIRS PRODUCED IN PARAMETRIC DOWN CONVERSION}}

\vspace{1cm}

{\bf Alberto Casado$^1$, Trevor Marshall$^2$, \\[0pt] Ram\'on Risco-Delgado$%
^1$ and Emilio Santos$^3$}.

$^1$ Escuela Superior de Ingenieros, Universidad de Sevilla,

41092 Sevilla, Spain.

$^2$Department of Mathematics, University of Manchester,

Manchester M13 9PL, U. K.

$^3$Departamento de F\'\i sica Moderna, Universidad de Cantabria,

39005 Santander, Spain.

\vspace{1cm}
\end{center}

\noindent {\bf Abstract}

In previous articles we have developed a theory of down conversion in
nonlinear crystals, based on the Wigner representation of the radiation
field. Taking advantage of the fact that the Wigner function is always
positive in parametric down conversion experiments, we construct a local
hidden variables model where the amplitudes of the field modes are taken as
random variables whose probability distribution is the Wigner function. In
order to achieve our goal we give a model of detection which is fully local
but departs from quantum theory. In our model the zeropoint (vacuum) level
of radiation lies below a threshold of the detectors and only signals above
the threshold are detectable. The predictions of the model agree with those
of quantum mechanics if the signal intensities surpase some level and the
efficiency is low. This is consistent with the known fact that quantum
mechanics is compatible with local realism in that case (a fact called the
``efficiency loophole''). Our model gives a number of constraints which do
not follow from the quantum theory of detection and are experimentally
testable.

PACS number(s): 42.50.Ar, 03.65.Sq, 42.50.Lc

\newpage

\section{Introduction}

\smallskip The aim of the present article is to propose a local hidden
variables (LHV) model for the experiments involving parametric down
conversion (PDC). The model is based on the quantum theory of PDC when
formulated in the Wigner representation. The reader may wonder what is the
use of such a LHV model and therefore this is the first question which we
should answer. Before doing that, we shall briefly comment on the present
status of the empirical tests of LHV theories.

As is well known, in 1964 Bell proved that there are predictions of quantum
mechanics which cannot be reproduced by any LHV theory \cite{Bell}.
Therefore it seems possible, in principle, to discriminate empirically
between quantum mechanics and ``local realism'' (i.e. the whole family of
LHV theories) by means of some specific experiments. Such experiments
usually attempt to test whether a Bell's inequality is violated and will be
called ``Bell's tests'' in the following. Actually most of the Bell's tests
made during the last 20 years have used photon pairs produced in the process
of PDC \cite{Rarity, Kwiat, weihs}. PDC has also become popular for the
study of other nonclassical aspects of light \cite{Horne, Franson} and, more
recently, for the development of the quantum theory of information like the
implementation of cryptographic schemes \cite{qcrip}, teleportation \cite
{Bennet}, etc.. In general the performed Bell's tests have confirmed quantum
mechanics but, in spite of the great effort made, they have been unable to
provide an uncontroversial disproof of local realism. The reason is that, up
to now, nobody has been able to test empirically a genuine Bell's
inequality, i.e. an inequality derived from the assumptions of realism and
locality alone. All inequalities actually tested in the performed
experiments involve additional assumptions, allegedly plausible, like {\em %
fair sampling}, {\em no-enhancement}, etc. \cite{Clauhorn}, \cite{Santos}.
The consequence is that {\em the important question whether LHV theories are
possible is still open}. This simple fact is rarely acknowledged in the
current literature. On the contrary, it has been repeatedly claimed, in
respected books and journals over the last thirty years, that local realism
has been refuted experimentally.

The essential difficulty for the performance of genuine Bell's tests with
optical photons has been the low efficiency of photon counters, which allows
for a loophole in the disproof of LHV theories known as the ``efficiency
loophole''. During the eighties and early nineties there was the hope that
the efficiency loophole could be easily blocked when detectors of high
efficiency (and low dark rate)\ were developed. In the last few years such
detectors have become available but nevertheless all attempts at blocking
the efficiency loophole have failed \cite{grangier}{\bf \ }and people are
turning to experiments not involving optical photons \cite{rowe}{\tt . }This
problem leads us naturally to ask why it is so difficult to block the
efficiency loophole in PDC experiments. One of the purposes of the present
article is to provide a partial resolution of this mystery. In fact, we
shall prove that it is possible to find a LHV model for all PDC experiments
by combining standard quantum theory for the production and propagation of
light with a non-quantum model of detection. Our detection model, although
not quantal, agrees with quantum theory for the performed experiments
(which, as said above, suffer from the detection loophole) but departs from
quantum mechanics in some (as yet unperformed) experiments, which might
therefore be able to discriminate between quantum mechanics and local
realism. In general, a LHV model for an experiment, or a class of
experiments, has two goals. Firstly it proves that LHV theories for those
experiments do actually exist, a proof which cannot be derived just from the
fulfillment of Bell's inequality because these are {\em necessary}, but not 
{\em sufficient}, conditions for local realism. Secondly LHV models play the
role of counterexamples. That is, their mere existence proves that the said
experiment is unreliable for the disproof of the whole family of LHV
theories. In the past, many LHV models for Bell's tests with PDC have
appeared in the literature \cite{yo y selleri, santospdc}. However, those
models were mathematical constructs without too much physical content and,
therefore, gave no hint as how to improve the experiments in order to make
them reliable. In contrast, the LHV model which we present in this article
gives a specific testable prediction: if that model (or a similar one) is
true, necessarily {\em there is a minimal light signal intensity which may
be reliably detected}. That minimal intensity depends on the geometry and
characteristics of the optical devices (lasers, nonlinear crystals, lenses,
etc.) and also on the properties of the photon counters used (in particular
their quantum efficiency). The model disagrees with quantum mechanics only
for specific combinations of detection efficiency and experimental set up,
and therefore determines the domain where a discrimination between quantum
mechanics and local realism may be possible. Consequently, the model
provides useful information about the domain where a reliable Bell's test,
using PDC photon pairs, may be performed. Such a test will not rely on the 
{\em plausibility }of some additional assumptions, as has been the case in 
{\em all} experiments performed up till now.

In addition, our model refutes the wisdom that the quantum zeropoint field
(the quantum vacuum fluctuations of the electromagnetic field) cannot be
real because in that case it would saturate all detectors. Indeed, there is
a long-standing controversy about whether the zeropoint field is real, or
merely an artifact of the quantization procedure; during the 1940s these
contrasting opinions were expressed by Casimir and Pauli respectively. If
the fluctuations are not real it is difficult to understand phenomena like
the Casimir effect or the Lamb shift. However, if they are considered real
two big problems arise. One of these is the huge gravitational effect that
it would produce at cosmic scales (more than 10$^{100}$ times greater than
the known upper limits on the cosmological constant). We shall not be
concerned with this problem here. The other difficulty is to understand how
a so big radiation (about 10$^5$ w/cm$^2$ in the visible spectrum alone)
does not blind photon detectors, so making it impossible to detect single
photon signals (as weak as 1 eV). Our model shows explicitly that the latter
difficulty may be overcome, if we assume that actual photon counters have a
detection threshold such that they detect only the radiation which surpasses
it. The threshold may be considered as arising from the fact that atoms (and
other quantum systems), in their ground state, are immersed in, and in
equilibrium with, the zeropoint radiation, so that only radiation above the
zeropoint level excites them. From this point of view, the quantum rule of
using normal ordering in the calculation of photon absorption probabilities
is seen as a mathematical procedure which takes account of the detection
threshold.

\section{The Wigner function and the local hidden variables}

According to Bell, the crucial difference between quantum mechanics and LHV
theories occurs in experiments where correlations between two particles at
space-like separation are measured (Einstein, Podolsky and Rosen, or EPR,
experiments \cite{epr}). Any LHV model should contain hidden variables $%
\lambda $, with a probability distribution $\rho (\lambda )$, giving the
following single and joint detection probabilities: 
\begin{equation}
\label{1}p_1=\int \rho (\lambda )P_1(\lambda ,\phi _1)d\lambda , 
\end{equation}

\begin{equation}
\label{2}p_2=\int \rho (\lambda )P_2(\lambda ,\phi _2)d\lambda , 
\end{equation}
\begin{equation}
\label{i1}p_{12}=\int \rho (\lambda )P_1(\lambda ,\phi _1)P_2(\lambda ,\phi
_2)d\lambda . 
\end{equation}
where $\phi _1$ and $\phi _2$ represent controllable parameters of the
experimental setup and $P_1(\lambda ,\phi _1)$, $P_2(\lambda ,\phi _2)$ are
some functions. If no further restrictions are put, a model resting upon Eqs.%
$\left( \ref{1}\right) $ to $\left( \ref{i1}\right) $ would be always
possible. However, in order to have a LHV model, the functions $P_1(\lambda
,\phi _1)$ and $P_2(\lambda ,\phi _2)$ should be probabilities and $\rho
(\lambda )$ a probability distribution; consequently the following
conditions should be fulfilled

\begin{equation}
\rho (\lambda )\geq 0\ \ \ \ ;\ \ \ \int \rho (\lambda )d\lambda =1. 
\end{equation}

\begin{equation}
\label{2a}0\leq P_{1}(\lambda ,\phi _{1}),\,P_{2}(\lambda ,\phi _{2})\leq 1. 
\end{equation}

\smallskip In some particular instances a method to construct an explicit
LHV model is to use the Wigner function. We consider a simple example. Let
us assume that we have two particles initially (time $t$=0) in a state
represented by the Wigner function W({\bf x}$_1$,{\bf p}$_1$,{\bf x}$_2$,%
{\bf p}$_2$). Now the particles evolve freely until time $t$ so that,
according to the quantum rules, the Wigner function becomes W(${\bf x}_1$-$%
{\bf p}_1t/m,{\bf p}_1,{\bf x}_2-{\bf p}_2t/m,{\bf p}_2$) (remember that the
quantum (Moyal) equation for the evolution of the Wigner function becomes
the classical (Liouville) equation in the case of free particles). Finally
we assume that at time $t$ we perform {\em local} measurements on the
particles such that we get the answer ``yes'' for the first (second)
particle if it lies inside a region R$_1$($\phi _1)$ (R$_2$($\phi _2)$). We
assume that the region R$_i$, $(i=1,2$ in the rest of the paper$)$ is
defined depending on one (or several) controllable parameters, $\phi _i,$ of
the experiment. In these conditions the probability of having the answer
``yes'' for both particles is 
\begin{equation}
\label{prob12}p_{12}=\int W({\bf x}_1-\frac{{\bf p}_1t}m,{\bf p}_1,{\bf x}_2-%
\frac{{\bf p}_2t}m,{\bf p}_2)\Theta [{\bf x}_1\in R_1(\phi _1)]\Theta [{\bf x%
}_1\in R_2(\phi _2)]d^3{\bf x}_1d^3{\bf x}_2,\smallskip \smallskip 
\smallskip
\end{equation}
where $\Theta [{\bf x}_i\in R_i]$ is the characteristic function of the
region $R_i,$ that is $\Theta $ $=$1 ($\Theta =$0) if ${\bf x}_i$ belongs
(does not belong) to the region $R_i.$ We see that the Wigner formalism
provides an explicit LHV model for the experiment {\em if the Wigner
function is positive. }In fact, in this case equation $\left( \ref{prob12}%
\right) ${\bf \ }is a particular case of equation $\left( \ref{i1}\right) $,
the initial positions and momenta $\left\{ {\bf x}_1,{\bf p}_1,{\bf x}_2,%
{\bf p}_2\right\} $ playing the role of the hidden variables $\lambda .$
Explicit use of the Wigner function in order to obtain a LHV model was made
by Bell himself \cite{bellwigner}. In that paper, Bell showed that the
situation proposed by Einstein, Podolsky and Rosen \cite{epr} had a positive
Wigner function, and hence he was able to obtain a LHV model in the form
outlined above. The problem is that only rarely the Wigner function of a
two-particle state is positive, so this method cannot be generalized.

If we pass from material particles to photons, then the Wigner function is
quite frequently positive definite. In particular this is the case for the
photon pairs produced in the PDC process (\cite{pdc1} to \cite{pdc6}). As it
is well known, in the Wigner formalism the operators of creation, $\hat a_{%
{\bf k}}^{\dagger }$ , and annihilation, $\hat a_{{\bf k}},$ of photons
become random complex amplitudes, $\alpha _{{\bf k}}^{*}$ and $\alpha _{{\bf %
k}}$, respectively. As the Wigner function is positive in this case, the
amplitudes may be interpreted as that of a real random electromagnetic
field, that is, these amplitudes become the local hidden variables, $\lambda
,$ of our model with the Wigner function playing the role of the function $%
\rho (\lambda )$ entering in Eqs.$\left( \ref{1}\right) $ to $\left( \ref{i1}%
\right) $. In the case of photons, however, the detection probability has
the form of a characteristic function, similar to $\Theta ({\bf x}_i\in R_i)$
above, only if we use the Glauber function commonly labelled P$\left(
\left\{ \alpha _{{\bf k}},\alpha _{{\bf k}}^{*}\right\} \right) $, but not
if we use the Wigner function. Indeed, in quantum optics, the P function is
assumed to correspond to the classical probability distribution of the
radiation amplitudes$.$ Therefore a LHV model parallel to that of Bell \cite
{bellwigner} should use the function P, rather than $W$, and it is indeed
well known that, whenever the P function is positive, the experiment does
admit a classical (LHV) interpretation.

If we insist on using the Wigner function, then it is not enough that it is
nonnegative in order to get a LHV model. In addition we must introduce some
non-negative functions of the random variables $\left\{ \alpha _{{\bf k}%
},\alpha _{{\bf k}}^{*}\right\} $, which play the role of the functions $%
P_i(\lambda ,\phi _i)$ of Eqs.$\left( \ref{1}\right) $ to $\left( \ref{i1}%
\right) .$ Those functions should fulfill two conditions: a) They should be
local, that is each function must depend only on the value of the
electromagnetic radiation entering the detector during a detection
time-window. b) They should give results so close to the quantum-optical
predictions that the model is compatible with all performed experiments.
Fulfilling these two conditions at the same time is not a trivial matter,
and it is the goal of the present paper to present such a function. So most
of the article will be devoted to exhibit a model of detection. Our LHV
model for the detection will rest upon the idea that the photodetector
performs some average of the electromagnetic field during the detection
window and within the detector volume, and that a photocount will be
produced when this averaged quantity exceeds a certain threshold value.

In summary, our previous work (\cite{pdc1} to \cite{pdc6}) has shown that
the Wigner function of the electromagnetic field at the detectors, derived
from radiation produced by PDC, is positive for all performed experiments,
and hence it provides an explicit LHV model for the {\em production} and 
{\em propagation} of this kind of radiation. The purpose of the present
paper is to develop a LHV model for the {\em detection}, therefore
completing an explicit LHV model for all PDC experiments.

In section 3 we briefly describe the generation and propagation of PDC light
in terms of the Wigner formalism. In section 4 we analyze the detection,
clarifying why even with a positive Wigner function it is not trivial to get
a LHV model. Section 5 is devoted to our detection model, the core of this
paper. In section 6 the explicit calculation of the probability distribution
of the LHVs of the model is done. In sections 7 and 8 we prove the agreement
of this LHV model with the quantum mechanical predictions in the low
efficiency limit. Finally in section 9 we discuss possible empirical tests
of our model versus quantum optics.

\section{Wigner formulation of PDC: generation and propagation of light}

In parametric down-conversion (PDC) an UV laser pumps a non-linear ($\chi ^2$%
) crystal and visible light is produced at a small angle (less than about 10$%
{{}^{\circ }}$) respect to the incoming laser beam. The relevant fact is
that there is a strong correlation between light beams produced at conjugate
directions. In particular if two detectors are placed in appropriate
positions in order to detect the PDC light, coincidence counts are seen with
a very high degree of temporal correlation (better than $10^{-13}$s). The
standard interpretation, resting upon the current quantum (Hilbert space)
analysis of the phenomenon, is that each photon of the laser ($\omega _0,${%
\boldmath $\,k$}$_0$) splits into two conjugated ones, ($\omega _1,${%
\boldmath $\,k$}$_1$) and ($\omega _2,${\boldmath $\,k$}$_2$), always
satisfying the matching conditions: $\omega _0=\omega _1+\omega _2$ and {%
\boldmath $k$}$_0\approx $ {\boldmath $k$}$_1+$ {\boldmath $k$}$_2$.
Conjugate beams have some interesting coherence properties which may be
interpreted considering that the two photons of a pair are {\em entangled}.
This is why PDC is considered a typically quantum phenomenon.

As is well known the Wigner function formalism may be used for the study of
quantum optics and it is completely equivalent to the more common, Hilbert
space formalism. Nevertheless the picture offered by the Wigner function is
quite different from the standard one. It emphasizes the wave aspects of
light whilst the Hilbert space formalism stresses the particle aspects
(photons are ``created'' in the source and ``annihilated'' at the
detectors). A well known property of the Wigner function is that it exhibits
explicitly the vacuum fluctuations of the electromagnetic field, also named
zeropoint field (ZPF). In the Wigner formalism, once we take the ZPF as
real, the phenomenon of PDC appears as a result of non-linear coupling,
inside the crystal, between the laser beam and the ZPF. A picture emerges
where everything looks like classical (Maxwell) wave optics, the Wigner
function being the probability distribution of the field amplitudes. Also
the electromagnetic radiation (including the ZPF) propagates according to
Maxwell theory. This provides a straightforward wave picture of the {\em %
generation} and {\em propagation} of light for all PDC experiments (\cite
{pdc1} to \cite{pdc6}). Crucial for this interpretation is that the Wigner
function is positive definite. In contrast the {\em detection} cannot be
interpreted classically in the Wigner formalism, at least not trivially. The
problem of detection will be considered in subsequent sections; here we
shall summarize the main features of generation and propagation within the
Wigner formalism.

In the Heisenberg picture of the Hilbert space formalism, the way to compute
single and joint detection probabilities (or rates) is based on the
following expressions (modulo some constants related to the efficiency of
the detectors):

\begin{equation}
\label{p1}p_i^q\propto \langle 0|\hat E^{-}(\mbox{{\bf r}}_i,t_i)\hat E^{+}(%
\mbox{{\bf
r}}_i,t_i)|0\rangle , 
\end{equation}

\begin{equation}
\label{p12}p_{12}^q\propto \langle 0|\hat E^{-}(\mbox{{\bf r}}_1,t_1)\hat
E^{-}(\mbox{{\bf r}}_2,t_2)\hat E^{+}(\mbox{{\bf r}}_2,t_2)\hat E^{+}(%
\mbox{{\bf
r}}_1,t_1)|0\rangle . 
\end{equation}
(We shall use the superscript $q$ when we refer to quantum mechanics and $m$
for our LHV model). $\hat E(\mbox{{\bf r}}_i,t_i)$ is the electric field
operator at the position and time $(\mbox{{\bf r}}_i,t_i)$, and, as usual,
we write $\hat E$ as a sum of positive and negative frequency parts, $\hat
E^{+}$ and $\hat E^{-}.$ These operators carry all the dynamical information
of the system because in the Heisenberg picture the state remain fixed
whilst the operators evolve in time. This is the reason why these
expectation values are computed with the initial (vacuum) state $|0\rangle $.

In PDC, when we apply the Wigner function formalism, the creation and
annihilation operators $\hat a_{{\bf k}}^{\dagger }$ and $\hat a_{{\bf k}}$
contained in $\hat E^{-}$ and $\hat E^{+}$ transform into random complex
(c-number) variables $\alpha _{{\bf k}}^{*}$ and $\alpha _{{\bf k}}$ (${\bf k%
}$ labels the normal modes of the electromagnetic field). Therefore, the
operators $\hat E^{-}$ and $\hat E^{+}$ transform into random functions $%
E^{-}$ and $E^{+}$ (which we represent without hat) of these variables. What
is the probability distribution for $\alpha _{{\bf k}}^{*}$ and $\alpha _{%
{\bf k}}$? Because we are working in the Heisenberg picture it may be seen (%
\cite{pdc1} to \cite{pdc6}) that it is the Wigner function for the {\em %
vacuum} state $|0\rangle $, that is, the Gaussian function 
\begin{equation}
\label{w}W(\{\alpha _{{\bf k}}^{*},\alpha _{{\bf k}}\})=\prod_{{\bf k}}\frac
2\pi e^{-2|\alpha _{{\bf k}}|^2}. 
\end{equation}
Then the electromagnetic field, now represented by the random functions $%
E^{-}$ and $E^{+}$ propagates in a totally classical way when passing
through all optical devices placed between the source and the detectors.

The image of PDC that we get when studying the process with the Wigner
formalism is similar to the classical one. Besides the laser, the crystal is
pumped by the vacuum field, its positive frequency part now given by (for
simplicity, we shall represent from now on the set of amplitudes $\{\alpha _{%
{\bf k}}^{*},\alpha _{{\bf k}}\}$ by $\alpha $):

\begin{equation}
\label{zpf}E_0^{(+)}({\bf r},t;\alpha )=i\sum_{{\bf k}}\left( \frac{\hbar
\omega }{\epsilon _0L_0^3}\right) ^{1/2}\alpha _{{\bf k}}e^{i{\bf k}\cdot 
{\bf r}-i\omega t}. 
\end{equation}
This means that there is not only one input field on the crystal and
therefore the PDC field is produced in a ``classical'' way as a consequence
of the non-linear coupling between the laser and the ZPF. The PDC field has
been computed in Refs. \cite{pdc5} and \cite{pdc6}, where it was shown the
influence of the size of the crystal and the radius of the pumping on the
matching conditions as well as the appearance of the typical rainbow in form
of cones. A simplified expression of this field is

\begin{equation}
\label{pdc}
\begin{array}{c}
E^{(+)}( 
{\bf r},t;\alpha )=i\sum_{{\bf k}}\left( \frac{\hbar \omega }{\epsilon
_0L_0^3}\right) ^{1/2}[\alpha _{{\bf k}}e^{i{\bf k}\cdot {\bf r}-i\omega
t}+g\alpha _{{\bf k}}^{*}e^{i({\bf k}_0-{\bf k)}\cdot {\bf r}-i(\omega
_0-\omega )t} \\ +\frac 12g^2\alpha _{{\bf k}}\text{ }e^{i{\bf k}\cdot {\bf r%
}-i\omega t}], 
\end{array}
\end{equation}
where three terms can be recognised. The first one is just the ZPF that
crosses the crystal without any modification. The second one is the new
field produced by the crystal as a consequence of the non-linear coupling
between the laser and the ZPF. $g$ is the coupling parameter which in
general depends on $\omega _0-\omega $; for the sake of clarity we shall not
write explicitly this dependence. The third term is similar to the first
one; it consists again of ZPF, but a little modified. It is necessary to
consider this term because the detection probability goes as $g^2$. In order
to make clear that we are dealing with a stochastic field, we have written
in $E^{^{(+)}}({\bf r},t;\alpha )$ the explicit dependence with the set of
random amplitudes, $\alpha $, which means that the PDC field has different
values for different realizations of the ZPF. The amplitudes $\alpha $
obviously represent the hidden variables of our model.

It is convenient to expand the PDC field in plane waves

\begin{equation}
\label{field}E^{(+)}({\bf r},t;\alpha )=\sum_{{\bf k}}\left( \frac{\hbar
\omega }{\epsilon _0L_0^3}\right) ^{1/2}\beta _{{\bf k}}\text{ }e^{-i{\bf k}%
\cdot {\bf r}+i\omega t}\text{,} 
\end{equation}
where the amplitudes $\beta _{{\bf k}}$ are linear functions of $\alpha $,
fulfilling the conditions \cite{pdc2}:

\begin{equation}
\label{genoveva}\left\langle \beta _{{\bf k}}\beta _{{\bf k}^{\prime
}}^{{}}\right\rangle =\left\langle \beta _{{\bf k}}^{*}\beta _{{\bf k}%
^{\prime }}^{*}\right\rangle =0\qquad ;\qquad \left\langle \beta _{{\bf k}%
}\beta _{{\bf k}^{\prime }}^{*}\right\rangle =0\ \ \ \ \ {\bf k\neq k}%
^{\prime }. 
\end{equation}

We end this section pointing out that, in the Wigner function formalism,
entanglement appears as a correlation between the ZPF associated to two
different light beams, whilst classical correlation just involves the signal
leaving the ZPF untouched \cite{pdc4}. This fact is becoming increasingly
clear even out of the context of the Wigner formalism \cite{ralph}

\section{The Wigner formulation of PDC: detection}

In order to complete the Wigner function approach to PDC experiments, it is
necessary to get the expression for the detection probability in terms of
the electromagnetic radiation arriving at the detector. For the sake of
clarity let us begin with the ideal case where the radiation field is
represented by a single mode. In this case, working in the Heisenberg
picture, the single detection probability is, according to quantum optics
(for simplicity we shall omit the subindex $i$ and refer to the detector $1$%
; obviously, the same expressions holds for the detector $2$):

\begin{equation}
\label{pw0}p_1^q\propto \langle \Phi |\hat b_{{\bf k}}^{\dagger }({\bf r}%
,t)\hat b_{{\bf k}}({\bf r},t)|\Phi \rangle =\frac 12\langle \Phi |\left[
\hat b_{{\bf k}}^{\dagger }({\bf r},t)\hat b_{{\bf k}}({\bf r},t)+\hat b_{%
{\bf k}}({\bf r},t)\hat b_{{\bf k}}^{\dagger }({\bf r},t)-1\right] |\Phi
\rangle 
\end{equation}
\begin{equation}
\label{pe}=\int W(\alpha )\left[ |\beta _{{\bf k}}({\bf r},t;\alpha
)|^2-\frac 12\right] d\alpha \equiv \langle |\beta _{{\bf k}}|^2-\frac
12\rangle _{_W}, 
\end{equation}
where $|\Phi \rangle $ is the initial state of the radiation, $W(\alpha )$
the corresponding Wigner function, and $\hat b_{{\bf k}}({\bf r},t)$ ( $\hat
b_{{\bf k}}^{\dagger }({\bf r},t)$ ) the time-dependent creation
(annihilation) operators, $\beta _{{\bf k}}$ and $\beta _{{\bf k}}^{*}$
being the corresponding amplitudes in the Wigner formalism. The first
equality derives from the use of the commutation relations and the second is
the passage from symmetrical ordered operators to the Wigner representation.
In the latter expression we have exhibited the dependence of the amplitude $%
\beta _{{\bf k}}$ on position and time, and on the initial amplitudes, $%
\alpha $. The simbol $\langle \rangle _{_W}$ means the average of the
quantity inside weighted with the Wigner function. In what follows we shall
omit the subindex $_W$.

In the general case, from (\ref{p1}) it is straightforward to get for a
point-like detector \cite{pdc1}:

\begin{equation}
\label{pw1}p_1^q({\bf r},t)\propto \int W(\alpha )\left[ I\left( {\bf r}%
,t;\alpha \right) -I_0\right] \ d\alpha =\langle I-I_0\rangle , 
\end{equation}
where $I({\bf r},t;\alpha )=c\varepsilon _0E^{+}({\bf r},t;\alpha )E^{-}(%
{\bf r},t;\alpha )$ is the intensity for a realization of the field (\ref
{field}) at the position and time $({\bf r},t)$, and $W(\alpha )$ is the
Wigner function of the initial state. $N$ is the number of modes (we should
take the limit $N\rightarrow \infty $ at some appropriate moment). $I_0$ is
the mean intensity of the ZPF, so that Eq.$\left( \ref{pw1}\right) $ might
be interpreted as stating that the detector has a threshold so that it only
detects the part of the field that is above the average ZPF. Two remarks are
in order: a) The intensity $I$ contains a (possibly complicated) dependence
on the initial amplitudes of all radiation modes, but $I_0$ is a constant
(compare with $\left( \ref{pe}\right) ).$ We might understand this fact by
saying that the detector ``knows'' the radiation actually arriving at a
given time, that is signal plus noise (ZPF), but it is not a trivial matter
to remove the noise. The quantum rule $\left( \ref{pw1}\right) $ is just to
subtract the mean, a formal procedure which cannot be physical because it
gives rise to ``negative probabilities'', as we shall discuss below. b)
Strictly the integral in $\left( \ref{pw1}\right) $ should involve all
radiation modes; therefore some cut-off frequency is required in order to
avoid divergences. Nevertheless, most of the radiation modes are usually not
``activated'' (in usual quantum language we say that they contain no
photons) and therefore they may be ignored. That is, the contribution of
these modes to the average $\langle I\rangle $ equals the contribution to $%
I_0$ so that ignoring them does not change the difference $\langle
I-I_0\rangle .$ We shall call ``relevant modes'' those which cannot be
ignored.

In the same way it is also possible to obtain the coincidence detection
probability for two detectors, placed at $({\bf r}_1,t_1)$ and $({\bf r}%
_2,t_2).$ We get from $\left( \ref{p12}\right) $

$$
\begin{array}{c}
p_{12}^q( {\bf r}_1,t_1;{\bf r}_2,t_2)\propto \int W(\alpha )\left[ I\left( 
{\bf r}_1,t_1;\alpha \right) -I_{10}\right] \left[ I\left( {\bf r}%
_2,t_2;\alpha \right) -I_{20}\right] d\alpha 
\end{array}
$$
\begin{equation}
\label{pw12}=\langle (I_1-I_{10})(I_2-I_{20})\rangle , 
\end{equation}
where we see that the coincidence detection probability is obtained by
multiplying the intensities that arrive at the two detectors, after
subtracting the threshold, and averaging this quantity with (\ref{w}) for
all the possible realizations of the field. Writing $I_{10}$ $\neq $ $I_{20}$
we emphasize that the thresholds of both detectors may be different (in
particular the relevant modes are usually different in the two cases).

The relevant question for us is whether expressions (\ref{pw1}) and (\ref
{pw12}) are suitable for a local realist interpretation. In fact, those
expressions look like Eqs.$\left( \ref{1}\right) $ to $\left( \ref{i1}%
\right) $ with $W(\alpha )$ playing the role of $\rho (\lambda )$ and $I-I_0$
the role of $P$. However, $I-I_0$ does not fulfil Eq.$\left( \ref{2a}\right) 
$; in particular it is not always positive and therefore the quantum theory
of detection prevents to get a trivial LHV theory from the Wigner
representation. The problem is not the huge value of the zeropoint energy
because the threshold intensity $I_0$ cancels precisely that intensity. The
problem lies in the fluctuation of the intensity. For the weak light signals
of the experiments the fluctuations of $I$ may be such that $I<I_0$.

In order to study whether this important problem may be solved we shall
begin showing that the removal of some idealities involved in Eqs. (\ref{pw1}%
) or (\ref{pw12}) alleviates the situation. Firstly these equations were
derived including only modes corresponding to a beam of almost parallel wave
vectors. If this is not the case we should write Eq. (\ref{pw12}) using the
Poynting vector rather than the intensity. The direction of Poynting vector
of the signal is well defined whilst that of the noise (the ZPF) is random
with zero mean, which may make easier the discrimination. More important is
the fact that the detection probability should not depend on the
instantaneous intensity at a point. In fact, it is natural to assume that
the detection probability depends on all the incoming radiation entering the
detector during the detection time window. We shall write Eqs. (\ref{pw1})
and (\ref{pw12}) in a more realistic form, with a different meaning for the
average represented by $\langle \rangle $. We have 
\begin{equation}
\label{641}p_1^q=\int W(\alpha )Q_1(\alpha ,\phi _1)d\alpha \;, 
\end{equation}
\begin{equation}
\label{64}p_{12}^q=\int W(\alpha )Q_1(\alpha ,\phi _1)Q_2(\alpha ,\phi
_2)d\alpha \;, 
\end{equation}
where 
$$
\label{65}Q_i(\alpha ,\phi _i)=\frac{\eta _i}{h\nu _i}\int_0^Tdt_i%
\int_Ad^2r_i\left[ I_i(\alpha ,\phi _i,{\bf r}_i,t_i)-I_{i0}\right] =\;\frac{%
\eta _i}{h\nu _i}\left[ \tilde I_i(\alpha ,\phi _i)-\tilde I_{i0}\right] , 
$$
\begin{equation}
\label{itilde}{} 
\end{equation}
where $\tilde I_i(\alpha ,\phi _i)-\tilde I_{i0}\equiv $ $\tilde
I_{is}(\alpha ,\phi _i)$ is the result of integrating the difference between
the actual intensity and the intensity of the zeropoint field over the time
window, $T$, and the surface aperture of the detector, $A$. We have divided
by the typical energy of one ``photon'' so that $Q_i$ becomes dimensionless.
In this way $\eta _i$ is the quantum efficiency of the detector.

The relevant question is whether (\ref{641}) and (\ref{64}) may now be
considered a particular case of (\ref{1}) to (\ref{i1}). If the answer is
affirmative (negative) the formalism provides (does not provide) an explicit
LHV model for the experiment. Actually the answer is not yet affirmative
because we cannot guarantee the positivity of $Q_i$. (For the additional
requirement, $Q_i\leq 1$, see below Eq. (\ref{5a}), but this condition
certainly holds for the low collection efficiencies of the experiments
performed up till now). The problem is now alleviated by the time and space
integrations in (\ref{itilde}). Indeed, the fluctuations of the intensity
are strongly reduced by averaging over space-time regions, as the Heisenberg
(uncertainty) relations show. But we can guarantee the positivity of $Q_i$
only in the limit of infinitely wide time-windows or infinitely large
apertures, which is non-physical. Consequently we conclude that it is not
possible to interpret directly the Wigner-function formalism as a LHV model
for the PDC experiments.

We shall devote the rest of the paper to describe a plausible model of
detector that ``works'' in a strictly local way.

\section{The detection model}

We will now proceed to show the basic points of our model:

\begin{enumerate}
\item  We shall assume that the detector is formed by a set of individual
photodetector elements, $D_j$, each characterized by a central frequency $%
\omega _j$, and a wave vector ${\bf k}_j$ ($\omega _j=|{\bf k}_j|/c$), to
which $D_j$ responds. We shall consider the direction of ${\bf k}_j$ to be
normal to the surface of detector, which is taken as a cylinder of area $%
A=\pi R^2$ and length $L$.

\item  If the light beam contains frequencies into the interval $(\omega
_{\min ,}\omega _{\max })$, then $\omega _j\in (\omega _{\min ,}\omega
_{\max }).$ Also, given two detector elements with frecuencies $\omega _j$
and $\omega _l$, the following relation holds

\begin{equation}
\label{nueva}
\begin{array}{c}
|\omega _j-\omega _l|\geq \frac{2\pi }T.
\end{array}
\end{equation}

\item  We shall suposse that the relevant quantity for the detection is not
directly the electromagnetic field $E^{(+)}({\bf r},t;\alpha )$, but a {\em %
filtered field} (by Fourier transform over the detector volume and time
window of detection). We shall define the filtered field corresponding to
detector element $D_j$, as 
\begin{equation}
\label{2a}\overline{E}_j^{(+)}(\alpha )=\frac 1{\pi
R^2LT}\int_VdV\int_0^TE^{(+)}({\bf r},t;\alpha )\mbox{e}^{-i{\bf k}_j\cdot 
{\bf r}+i\omega _jt}dt.
\end{equation}
This equation shows that $\overline{E}_j^{(+)}$ will depend only on the
radiation that crosses the detector during a time window, which is the {\em %
locality constraint }on the model.

By substituting Eq. (\ref{field}) into Eq. (\ref{2a}) and taking the origin
of the reference system $OXYZ$ at the center of the crystal with the $OZ-$%
axis as its axis, we obtain, after some easy algebra 
\begin{equation}
\label{5}\overline{E}_j^{(+)}(\alpha )=\sum_{{\bf k}}\left( \frac{\hbar
\omega }{\epsilon _0L_0^3}\right) ^{1/2}\beta _{{\bf k}}\frac{2J_1(k_rR)}{%
k_rR}\mbox{sinc}\left[ \frac L2(k_z-\frac{\omega _j}c)\right] \mbox{sinc}%
\left[ \frac T2(\omega -\omega _j)\right] .
\end{equation}

The presence of the \mbox{sinc} factor in Eq. (\ref{5}) implies that, for a
given value of $\omega _j$, only frequencies whithin a range of width $2\pi
/T$ centered at $\omega _j$ will contribute to the sum in (\ref{5}), i.e.
the photodetector element $Dj$ is sensitive to radiation with frequencies in
the interval $(\omega _j-\frac{\Delta \omega }2,\omega _j+\frac{\Delta
\omega }2)$ with $\Delta \omega \approx 2\pi /T.$

\item  The maximun number of independent detecting elements is (see Eq. (\ref
{nueva})) 
\begin{equation}
\label{i5}N\approx \frac{\delta \omega }{\Delta \omega }\approx \frac T\tau ,
\end{equation}

where $\delta \omega =\omega _{\max }-\omega _{\min }\approx 2\pi /\tau ,$ $%
\tau $ being the coherence time of beam. Typical values for $T\ (\approx
10^{-8}\,s)$ and $\tau \ (\approx 10^{-12}\,s)$ give rise to $N=10^4$.

Note that the modes which contribute to $\overline{E}_j^{(+)}$ are different
to those corresponding to $\overline{E}_k^{(-)}$, which implies that $%
\overline{E}_j^{(+)}$ and $\overline{E}_k^{(-)}$ are uncorrelated for $j\neq
k$ (see Eq.(\ref{genoveva})) : 
\begin{equation}
\label{6}\langle \overline{E}_j^{(+)}\overline{E}_k^{(-)}\rangle \approx
0\;\ \ ;\ \ \ j\neq k.
\end{equation}

This expression will be useful later on.

\item  Now we shall define the {\em effective intensity, }$\overline{I}%
(\alpha ),$ which is obtained from the filtered fields $\left( \ref{2a}%
\right) $ in the form

\begin{equation}
\label{imed}\overline{I}(\alpha )=c\epsilon _0\sum_{j=1}^N\overline{E}%
_j^{(+)}(\alpha )\overline{E}_j^{(-)}(\alpha ),
\end{equation}
and replace Eq.(\ref{itilde}) by the expression 
\begin{equation}
\label{noexp}P(\alpha ,\phi )=\zeta (\overline{I}(\alpha ,\phi )-\overline{I}%
_0)\Theta [\overline{I}(\alpha ,\phi )-I_m],
\end{equation}
where $\overline{I}_0$ is the average of $\overline{I}$ for the ZPF. $I_m$
is some threshold intensity related to the ``voltaje biass'' of the
detector, fulfilling the condition $I_m>\overline{I}_0,$ and $\Theta (x)$ is
the Heavside function $\Theta (x)$ $=1$ if $x>0$, $0$ otherwise. We have
considered all the complex dependence of the intensity on the initial
amplitudes, $\alpha $, and the controllable parameters in the experiment, $%
\phi $.

We see that taking $\zeta \rightarrow \eta TA/h\nu ,$ our Eq.$\left( \ref
{noexp}\right) $ becomes the standard quantum detection probability Eq.$%
\left( \ref{pw1}\right) $ if we consider instantaneous detection (i.e. $%
T\rightarrow 0)$ by a point-like detector (i.e. $R\rightarrow 0,L\rightarrow
0)$ and remove the Heavside function. We would like to interpret Eq.(\ref
{noexp}) as the probability for a given realization of the field and it is
analogous to $P_1(\lambda ,\phi _1)$ or $P_2(\lambda ,\phi _2)$ in Eqs. (\ref
{1}) and (\ref{2}) for a value of the hidden variable. However, although the
Heavside function ensures that the quantity $P(\alpha ,\phi )$ in (\ref
{noexp}) is positive, it does not fulfil the condition $P(\alpha ,\phi )\leq
1$ in general. Therefore we should propose an expression which is positive,
lower than $1,$ and reduces to Eq. (\ref{noexp}) in the case of $\zeta |%
\overline{I}(\alpha ,\phi )-\overline{I}_0|\ll 1$. A simple expression
achieving these goals is 
\begin{equation}
\label{5a}P(\alpha ,\phi )=(1-\mbox{e}^{-\zeta (\overline{I}(\alpha ,\phi )-%
\overline{I}_0)})\Theta [\overline{I}(\alpha ,\phi )-I_m],
\end{equation}
which completes the definition of our model.

\item  The single detection probability predicted by the model is

\begin{equation}
\label{lorena}p_1^m=\int W(\alpha )P(\alpha ,\phi )d\alpha ,
\end{equation}

which can be expresed in the following equivalent form: 
\begin{equation}
\label{dete}p_1^m=\int \rho (\overline{I})P(\overline{I})\ d\overline{I}\ ,
\end{equation}

where 
\begin{equation}
\label{dete11}\rho (\overline{I})=\int W(\alpha )\delta [\overline{I}%
-c\epsilon _0\sum_{j=1}^N\overline{E}_j^{(+)}(\alpha ,\phi )\overline{E}%
_j^{(-)}(\alpha ,\phi )]d\alpha .
\end{equation}
Similarly we have, for the joint detection probability,

\begin{equation}
\label{dete2}p_{12}^m=\int \rho _{12}(\overline{I}_1,\overline{I}_2)P_1(%
\overline{I}_1)P_2(\overline{I}_2)d\overline{I}_1d\overline{I}_2,
\end{equation}
\end{enumerate}

where 
\begin{equation}
\label{dete22}
\begin{array}{c}
\rho _{12}( 
\overline{I}_1,\overline{I}_2)=\int W(\alpha )\delta [\overline{I}%
_1-c\epsilon _0\sum_{j=1}^N\overline{E}_{j1}^{(+)}(\alpha ,\phi _1)\overline{%
E}_{j1}^{(-)}(\alpha ,\phi _1)] \\ \times \delta [\overline{I}_2-c\epsilon
_0\sum_{j=1}^N\overline{E}_{j2}^{(+)}(\alpha ,\phi _2)\overline{E}%
_{j2}^{(-)}(\alpha ,\phi _2)]d\alpha . 
\end{array}
\end{equation}
Eqs. (\ref{dete}) and (\ref{dete2}) are very convenient for the comparison
between our model and quantum optics. As said above, the amplitudes $\alpha $
are the hidden variables, $W(\alpha )$ playing the role of the function $%
\rho (\lambda )$ in Bell's formulation. The analogous of the probabilities $%
P $($\lambda ,\phi )$ are 
\begin{equation}
\label{Qj}P(\alpha ,\phi )=\int P(\overline{I})\delta [\overline{I}%
-c\epsilon _0\sum_{j=1}^N\overline{E}_j^{(+)}(\alpha ,\phi )\overline{E}%
_j^{(-)}(\alpha ,\phi )]\ d\overline{I}. 
\end{equation}
The (complicated) dependence of the filtered field on the vacuum amplitudes
should be derived from the quantum Wigner formalism for every experiment.
This was made in Refs. \cite{pdc1} to \cite{pdc4}.

Now we shall study whether the predictions of our LHV model are compatible
with the results of performed experiments and to which extent they agree
with the quantum predictions. For that purpose the following steps will be
taken: {\em i)} To obtain the probability distribution, $\rho _0(\overline{I}%
),$ of the effective intensity of the ZPF in absence of any further
electromagnetic radiation. {\em ii)} To obtain the probability distribution, 
$\rho (\overline{I}),$ for the radiation that arrives at the detector when
it is illuminatated with the light coming from a PDC process. {\em iii)} To
compute single detection probabilities and {\em iv)} joint detection
probabilities.

\section{Calculation of $\rho _0(\overline{I})$ and $\rho (\overline{I})$}

The filtered field $\overline{E}_j^{(+)}$ is Gaussian because it derives
from a Gaussian random field, $E_j^{+}$, under linear transformations.
Consequently $\overline{E}_j^{(+)}\overline{E}_j^{(-)}$ is an exponential
random variable. Nevertheless, by virtue of $\overline{I}$ being the sum of
a large number of independent random variables $(\approx 10^4)$, a version
of the Central Limit Theorem applies, so that $\overline{I}$ is Gaussian to
a good approximation. Consequently the full probability distribution is
determined by the mean and the standard deviation.

For instance, the probability distribution for the effective intensity when
only the zeropoint field is present is 
\begin{equation}
\label{8}\rho _0(\overline{I})=\frac 1{\sqrt{2\pi }\sigma _0}\mbox{e}^{\frac{%
(\overline{I}-\overline{I}_0)^2}{2\sigma _0^2}}. 
\end{equation}

Let us calculate $\overline{I}_0$ and $\sigma _0$. By using Eqs. (\ref{5})
(making $\beta _{{\bf k}}\rightarrow \alpha _{{\bf k}})$ and (\ref{imed}),
we have

\begin{equation}
\label{suma}\overline{I}_0=\sum_{j=0}^N\overline{I}_{0j}, 
\end{equation}
where

\begin{equation}
\label{9}\overline{I}_{0j}=\frac \hbar {L_0^3}\sum_{{\bf k}}\frac c2\frac{%
4J_1^2(k_rR)}{(k_rR)^2}\mbox{sinc}^2\left[ \frac L2(k_z-\frac{\omega _j}%
c)\right] \mbox{sinc}^2\left[ \frac T2(\omega -\omega _j)\right] . 
\end{equation}
Here we have made use of the relation $\langle \alpha _{{\bf k}}\alpha _{%
{\bf k}^{\prime }}^{*}\rangle =\delta _{{\bf kk}^{\prime }}/2,$ which can be
easily derived from Eq. (\ref{w}).

Taking the continuous limit $\sum_{{\bf k}}/L_0^3\rightarrow \int d^3{k}%
/(2\pi )^3$, and changing to spherical polar coordinates $%
(k_x,k_y,k_z)\,\rightarrow (\omega ,\theta ,\psi )$: 
$$
k_z=\frac \omega c\mbox{cos}\theta \,\,\,\,;\,\,\,\,k_x=\frac \omega c\,%
\mbox{sen}\theta \,\mbox{cos}\psi \,\,\,\,;\,\,\,\,k_y=\frac \omega c\,%
\mbox{sen}\theta \,\mbox{sen}\psi , 
$$
we get 
$$
\overline{I}_{0j}=\frac{2\pi c\hbar }{2(2\pi )^3}\int_0^\infty \frac{\omega
^2}{c^2}\frac{d\omega }c\mbox{sinc}^2\left[ \frac T2(\omega -\omega
_j)\right] 
$$
\begin{equation}
\label{10}\times \int_0^{\pi /2}d\theta \ \mbox{sin}\theta \frac{%
4J_1^2(\frac \omega c\mbox{sin}\theta R)}{(\frac \omega c\mbox{sin}\theta
R)^2}\mbox{sinc}^2\left[ \frac L{2c}(\omega \mbox{cos}\theta -\omega
_j)\right] . 
\end{equation}
The typical value of $T\omega $, $\omega $ being a frequency in the visible
range, is $10^7.$ For this reason we can approximate the first sinc factor
in the above expression by a delta function. After performing the
integration in $\omega $, we obtain: 
\begin{equation}
\label{11}\overline{I}_{0j}=\frac{\hbar \omega _j^3}{4c^2T\pi }\int_0^{\pi
/2}d\theta \ \mbox{sin}\theta \frac{4J_1^2(\frac{\omega _j}c\mbox{sin}\theta
R)}{(\frac{\omega _j}c\mbox{sin}\theta R)^2}\mbox{sinc}^2\left[ \frac{%
L\omega _j}{2c}(\mbox{cos}\theta -1)\right] . 
\end{equation}
Now, making the substitution $v=L\omega _j(1-\cos \theta )/2c$, we have

\begin{equation}
\label{11a}\overline{I}_{0j}=\frac{\hbar \omega _j^2}{2\pi cTL}%
\int_0^{L\omega _j/(2c)}dv\,\mbox{sinc}^2(v)\,\frac{4J_1^2\left[ \frac{%
2\omega _jR}c\sqrt{\frac{cv}{L\omega _j}(1-\frac{cv}{L\omega _j})}\right] }{%
\left[ \frac{2\omega _jR}c\sqrt{\frac{cv}{L\omega _j}(1-\frac{cv}{L\omega _j}%
)}\right] ^2}. 
\end{equation}
In this expression the upper limit of the integral is much larger than $1,$
because $L\gg \lambda $, $\lambda $ being the typical wavelenght of the
radiation. For this value of $v$ the sinc$^2$ factor is negligable. Thus we
can safely extend the upper limit to infinity. Also, the fraction $%
4J_1^2(x)/x^2$ can be approximated by $1$ in case that $x^2/4\ll 1$. It can
be shown that this implies the condition $R<\sqrt{\lambda L/8\pi ^2},$ a
situation that stands for usual small detector radius. Under these
conditions, we obtain

\begin{equation}
\overline{I}_{0j}=\frac{\hbar \omega _j^2}{4cTL}.
\end{equation}
From (\ref{suma}), the total mean intensity is

$$
\overline{I}_0=\sum_{j=1}^N\frac{\hbar \omega _j^2}{4cTL}=\sum_{j=0}^{T%
\delta \omega /(2\pi )}\frac{\omega _{min}+2\pi j/T}{4cTL}. 
$$
\begin{equation}
{} 
\end{equation}
Passing to the continuous, we finally obtain

\begin{equation}
\label{icero}\overline{I}_0=\frac T{2\pi }\int_{\omega _{min}}^{\omega
_{max}}\frac{\hbar \omega _j^2}{4cTL}=\frac{\hbar \overline{\omega }^2\delta
\omega }{8\pi cL}, 
\end{equation}
where $\overline{\omega }=(\omega _{max}+\omega _{min})/2.$

Let us now proceed to the calculation of $\sigma _0$. We have

\begin{equation}
\sigma _0^2=\langle \overline{I}^2\rangle -\overline{I}_0^2=c^2\epsilon
_0^2\sum_j\sum_l\left[ \langle \overline{E}_{0j}^{(+)}\overline{E}_{0j}^{(-)}%
\overline{E}_{0l}^{(+)}\overline{E}_{0l}^{(-)}\rangle -\langle \overline{E}%
_{0j}^{(+)}\overline{E}_{0j}^{(-)}\rangle \langle \overline{E}_{0l}^{(+)}%
\overline{E}_{0l}^{(-)}\rangle \right] . 
\end{equation}
Taking into account that we are dealing with Gaussian processes and that $%
\langle \overline{E}_{0j}^{(+)}\overline{E}_{0l}^{(+)}\rangle =\langle 
\overline{E}_{0j}^{(-)}\overline{E}_{0l}^{(-)}\rangle =0$ $(\forall \ j,$ $%
l) $, and $\langle \overline{E}_{0j}^{(+)}\overline{E}_{0l}^{(-)}\rangle
\approx 0$ if $l$ $\neq j$ (this follows from the fact that we are dealing
with a set of $N$ practically independent stochastic variables), we find

\begin{equation}
\label{sigmai}\sigma _0^2\approx \sum_j\overline{I}_{0j}{}^2\approx N^{-1}%
\overline{I}_0{}^2. 
\end{equation}
Hence we obtain

\begin{equation}
\label{sigmacero}\sigma _0=\sigma _{0j}\sqrt{N}=\langle \overline{I}%
_0\rangle \sqrt{\frac \tau T}, 
\end{equation}
and the final expression for $\rho _0(\overline{I})$ is

\begin{equation}
\label{42}\rho _0(\overline{I})=\frac{\sqrt{\frac T\tau }}{\sqrt{2\pi }\ 
\overline{I}_0}\mbox{e}^{-\frac{(\overline{I}-\overline{I}_0)^2T}{2\overline{%
I}_0{}^2\tau }}, 
\end{equation}
where $\overline{I}_0$ is given in Eq. $\left( \ref{icero}\right) .$

Let us proceed to the calculation of the probability distribution for the
effective intensity when there is a PDC signal present, $\rho (\overline{I}%
). $ Such distribution, as that of $\rho _0(\overline{I}),$ is also a
Gaussian. The argument is parallel to that leading to Eq.$\left( \ref{42}%
\right) $ starting from the fact that the filtered field (see Eq. $\left( 
\ref{5}\right) )$ is Gaussian and its statistical properties are close to
those of the ZPF, except for the greater intensity of the former for some
modes of the radiation. The Gaussian character of the PDC radiation has been
discussed in detail in our previous papers ($\cite{pdc1}$ to $\cite{pdc6}).$
But there is also another argument derived from the fact that the counting
statistics corresponds to a Gaussian (Glauber) P function (see \cite{teisch}%
). Now, it is known that the Wigner function is the convolution of the P
function with the Wigner function of the vacuum, which leads to a final
Gaussian Wigner function. The mentioned convolution has an interesting
physical interpretation: It corresponds to a random variable which is the
sum of two uncorrelated random variables. It is as if the signal is
superimposed incoherently to the ZPF when P$\geq 0,$ a situation usually
named ``classical light''.

The relation between the mean and the standard deviation is completely
analogous to that of the ZPF alone, and the distribution may be written

\begin{equation}
\label{rodei}\rho (\overline{I})=\frac{\sqrt{\frac T\tau }}{2\pi \overline{I}%
_0}\mbox{e}^{-\frac{(\overline{I}-\langle \overline{I}\rangle )^2T}{2\langle 
\overline{I}\rangle ^2\tau }}\cong \frac{\sqrt{\frac T\tau }}{\sqrt{2\pi }\ 
\overline{I}_0}\mbox{e}^{-\frac{(\overline{I}-\overline{I}_0-\overline{I}%
_s)^2T}{2\overline{I}_0{}^2\tau }} 
\end{equation}
where we have defined the signal mean effective intensity by 
\begin{equation}
\overline{I}_s=\langle \overline{I}\rangle -\overline{I}_0, 
\end{equation}
and taken into account that $\overline{I}_s\ll \overline{I}_0$ in the
denominator of the exponent in (\ref{rodei}). This can be easily
demonstrated for typical values of $L,$ $\tau ,$ and $\lambda $ in Eq. (\ref
{icero}), and by considering the intensity arriving at the detection system
in a usual PDC experiment.

\section{The single detection probability}

\label{secsingle}

We want to start by comparing the prediction of our model with that of
quantum optics. The quantum optical prediction for the single detection
probability is given by Eq. (\ref{641}). By using the properties of the
delta function this expression can be written in a different way:

$$
p_1^q=\frac \eta {h\nu }\int_0^\infty d\tilde I\ (\tilde I-\tilde I_0)\int
d\alpha \,W(\alpha )\delta [\tilde I(\alpha ,\phi )-\tilde I] 
$$
\begin{equation}
=\frac \eta {h\nu }\int_0^\infty \rho (\tilde I)(\tilde I-\tilde I_0)\
d\tilde I=\frac \eta {h\nu }\langle \tilde I_s\rangle , 
\end{equation}
where

\begin{equation}
\rho (\tilde I)=\int d\alpha \,W(\alpha )\delta [\tilde I(\alpha ,\phi
)-\tilde I]. 
\end{equation}

Now let us analyze the predictions of our model. They are given by Eq. (\ref
{dete}) where $P(\overline{I})$ was defined in Eq.(\ref{5a}) and $\rho (%
\overline{I})$ in Eq.(\ref{dete11}$),$ the latter calculated in section 6.
After some easy algebra, we obtain:

\begin{equation}
\label{hola}p_1^m=\frac 12\mbox{erfc}[\frac{I_m-\overline{I}_0}{\sigma _0%
\sqrt{2}}-\frac{\overline{I}_s}{\sigma _0\sqrt{2}}]-\frac 12\mbox{e}^{-%
\overline{I}_s\zeta }\mbox{e}^{\frac{\zeta ^2\sigma _0^2}2}\mbox{erfc}[\frac{%
I_m-\overline{I}_0}{\sigma _0\sqrt{2}}-\frac{\overline{I}_s}{\sigma _0\sqrt{2%
}}+\frac{\zeta \sigma _0}{\sqrt{2}}]. 
\end{equation}

In the above expression the important parameters $\overline{I}_s/\sigma _0,$ 
$(I_m-\overline{I}_0)/\sigma _0$, and $\overline{I}_s\zeta $ enter. Let us
now consider the two following situations: {\em i)} $\zeta \bar I_s\ll 1$
and $\zeta \sigma _0\ll 1$(linear approximation), and {\em ii)} $\zeta \bar
I_s\gg 1.$

{\em i) $\zeta \bar I_s\ll 1$ }and{\em \ $\zeta \sigma _0\ll 1$}$.$ This
should be the normal situation in experimental practice. In order to compute
the detection probability in this case, we take into account that the
relevant values of $\overline{I}-\overline{I}_0$ in the integral of Eq. (\ref
{dete}) are close to $\overline{I}_s$, which allows to make the
approximation $1-\mbox{e}^{-\zeta (\overline{I}-\overline{I}_0)}\cong \zeta
\left( \overline{I}-\overline{I}_0\right) .$ We obtain

\begin{equation}
\label{hh}p_1^m=\frac{{\em \zeta \bar I_s}}2\mbox{erfc}[\frac{I_m-\overline{I%
}_0-\bar I_s}{\sigma _0\sqrt{2}}]+\frac{\zeta \sigma _0}{\sqrt{2\pi }}%
\mbox{e}^{-\frac{(I_m-\overline{I}_0-\bar I_s)^2}{2\sigma _0^2}}. 
\end{equation}

Now, we may choose $I_m$ so that 
\begin{equation}
\label{cond2}\overline{I}_s-(I_m-\overline{I}_0)\gg \sigma _0, 
\end{equation}
With these two approximations we arrive to the following result:

\begin{equation}
\label{P1M}p_1^m\cong \varsigma \bar I_s. 
\end{equation}
The important point is that this result is very similar to the quantum one,
the only difference being the presence of the effective intensity of the
signal, $\bar I_s$, instead of the actual intensity, $\langle \widetilde{I}%
_s\rangle /AT$, in the quantum formula. In experimental practice a lens is
placed in front of the detector in such a way that the signal field has
spatial coherence on the surface of the lens. The condition for having
spatial coherence is 
\begin{equation}
\label{spc}d\lambda \geq R_lR_C, 
\end{equation}
$d$ being the typical distance between the nonlinear medium (with an active
radius $R_C$) and the detector; $R_l$ is the radius of the lens. It can be
demonstrated (see Appendix) that this property of the signal field implies
that 
\begin{equation}
\label{noll}\bar I_s=\frac{\langle \widetilde{I}_s\rangle }{AT}. 
\end{equation}
Hence, by substituting Eq. (\ref{noll}) into Eq. (\ref{P1M}), we have

$$
p_1^m\cong \varsigma \bar I_s=\frac \eta {h\nu }\langle \widetilde{I}%
_s\rangle , 
$$
a result that coincides with Eq. (\ref{641}).

If $\overline{I}_s=0$ strictly, in sharp contrast with quantum optics, our
model predicts the existence of some counts in any detector even in the
absence of signal. In fact, from Eq. (\ref{hh}) we get

\begin{equation}
\label{dark}p_{dark}^m=\frac{\zeta \sigma _0}{\sqrt{2\pi }}\mbox{e}^{-\frac{%
(I_m-\overline{I}_0)^2}{2\sigma _0^2}}, 
\end{equation}
which is very small if the condition 
\begin{equation}
\label{cond1}I_m-\overline{I}_0\gg \sigma _0, 
\end{equation}
is fulfilled. In this case there is no conflict with experiments, because we
may interpret the counts without signal as a part of the dark rate of the
detector.

{\em ii) }$\zeta \bar I_s\gg 1$. By taking this limit in Eq. (\ref{dete})
one can easily check that under this condition $p_1^m=1$ (provided that the
constraint (\ref{cond2}) holds true). That is, the detector saturates when
the intensity is very high, and gives a count in every time window, a fact
not in disagreement with experiments (or with quantum-optical predictions).

\smallskip\ \smallskip\ However, the remarkable agreement with
quantum-optical predictions is not enough to guarantee that we have arrived
to a LHV model for the detection probability. This is because in our model
there are two constraints that must be fulfilled: the constraint (\ref{cond2}%
), required for the linearity of the response at low efficiency, Eq.(\ref
{P1M})$,$ and (\ref{cond1}), needed for the smallness of the dark counting
probability, Eq.(\ref{dark})$.$ For these two conditions to be fulfilled it
is necessary that $\bar I_s\gg \sigma _0,$ a condition which makes difficult
to construct a LHV model for detection and, at the same time, what gives
predictive power to the model, as will be discussed in detail in the final
section.

\section{The joint detection probability}

The quantum mechanical prediction for the joint detection probability is
given by Eq. (\ref{64}). As before, using the properties of the delta
function this expression can be written in the following way:

$$
\begin{array}{c}
p_{12}^q=\int_0^\infty d\tilde I_1\int_0^\infty d\tilde I_2\int d\alpha
\,W(\alpha )\delta [\tilde I_1(\alpha ,\phi _1)-\tilde I_1]\delta [\tilde
I_2(\alpha ,\phi _2)-\tilde I_2] \\ 
\times \frac{\eta _1}{h\nu _1}(\tilde I_1-\tilde I_{10})\frac{\eta _2}{h\nu
_2}(\tilde I_2-\tilde I_{20}) 
\end{array}
$$
\begin{equation}
=\int_0^\infty d\tilde I_1\int_0^\infty d\tilde I_2\rho (\tilde I_1,\tilde
I_2)\frac{\eta _1}{h\nu _1}(\tilde I_1-\tilde I_{10})\frac{\eta _2}{h\nu _2}%
(\tilde I_2-\tilde I_{20})=\frac{\eta _1\eta _2}{h^2\nu _1\nu _2}\langle
\tilde I_{1s}\tilde I_{2s}\rangle , 
\end{equation}
where

\begin{equation}
\rho (\tilde I_1,\tilde I_2)=\int d\alpha \,W(\alpha )\delta [\tilde
I_1(\alpha ,\phi _1)-\tilde I_1]\delta [\tilde I_2(\alpha ,\phi _2)-\tilde
I_2]. 
\end{equation}

Now let us analyze the predictions of our model for the joint detection
probability. They are given by Eq. (\ref{dete2}) where $P_i(\overline{I}_i)$
was defined in Eq. (\ref{5a}) and $\rho (\tilde I_1,\tilde I_2)$ in Eq. (\ref
{dete22}):

$$
p_{12}^m=\int_{I_{1m}}^\infty \int_{I_{2m}}^\infty \rho _{12}(\overline{I}_1,%
\overline{I}_2)(1-\mbox{e}^{-\zeta _1(\overline{I}_1-\overline{I}_{10})})(1-%
\mbox{e}^{-\zeta _2(\overline{I}_2-\overline{I}_{20})})d\overline{I}_1d%
\overline{I}_2. 
$$

Because $\rho (\overline{I}_1,\overline{I}_2)$ is a doble Gaussian function,
it is defined by the mean values of its marginals, their standard deviations
and the correlation function $\langle (\overline{I}_1-\overline{I}_{1s}-%
\overline{I}_{10})(\overline{I}_2-\overline{I}_{2s}-\overline{I}%
_{20})\rangle $. In the case that $\sigma _1=\sigma _2=\sigma _{0,}$ then

\begin{equation}
\rho (\overline{I}_1,\overline{I}_2)=(2\pi \sigma _0^2)^{-1}\left( 1-\frac{%
\langle \hat I_1\hat I_2\rangle ^2}{\sigma _0^4}\right) ^{-1/2}\mbox{exp}%
\left[ -\frac{\hat I_1^2+\hat I_2^2-2\hat I_1\hat I_2\langle \hat I_1\hat
I_2\rangle /\sigma _0^2}{2(\sigma _0^2-\frac{\langle \hat I_1\hat I_2\rangle
^2}{\sigma _0^2})}\right] , 
\end{equation}
where we have defined $\hat I_1=\overline{I}_1-\overline{I}_{1s}-\overline{I}%
_{10}$ and $\hat I_2=\overline{I}_2-\overline{I}_{2s}-\overline{I}_{20}$.

Now, by considering the limits $\zeta _i\overline{I}_{is}\ll 1$ and $%
\overline{I}_{is}-(I_{im}-\overline{I}_{i0})\gg \sigma _0$, we arrive to the
following result:

$$
\begin{array}{c}
p_{12}^m\simeq \int_0^\infty d 
\overline{I}_1\int_0^\infty d\overline{I}_2\rho (\overline{I}_1,\overline{I}%
_2)\zeta _1(\overline{I}_1-\overline{I}_{10})\zeta _2(\overline{I}_2-%
\overline{I}_{20}) \\ =\zeta _1\zeta _2\langle (\overline{I}_1-\overline{I}%
_{20})(\overline{I}_1-\overline{I}_{20})\rangle , 
\end{array}
$$
\begin{equation}
{} 
\end{equation}
a result very similar to the quantum prediction, Eq. (\ref{64}).

The parameters in $\rho(\overline I_1,\overline I_2)$ are easily obtained
from the marginals and the correlation function by making equal their values
to the quantum mechanical predictions:

\begin{equation}
\int_0^\infty d\overline{I}_2\int_0^\infty \overline{I}_1\rho (\overline{I}%
_1,\overline{I}_2)\,d\overline{I}_1=\int_0^\infty d\overline{I}_1\rho (%
\overline{I}_1)\overline{I}_1\equiv \overline{I}_{10}+\overline{I}_{1s}, 
\end{equation}

\begin{equation}
\int_0^\infty d\overline{I}_2\int_0^\infty \overline{I}_2\rho (\overline{I}%
_1,\overline{I}_2)\,d\overline{I}_1=\int_0^\infty d\overline{I}_2\rho (%
\overline{I}_1)\overline{I}_2\equiv \overline{I}_{20}+\overline{I}_{2s}, 
\end{equation}

\begin{equation}
\int_0^\infty d\overline{I}_2\int_0^\infty d\overline{I}_1\overline{I}_1%
\overline{I}_2\rho (\overline{I}_1,\overline{I}_2)\equiv \langle \overline{I}%
_1\overline{I}_2\rangle . 
\end{equation}

\section{Discussion: Empirical tests of the model}

The predictions of our LHV model agree with those of quantum theory for all
PDC experiments with low efficiency, as stated above. Therefore our model
violates all ``Bell's inequalities'' empirically tested up to now, those
inequalities having been derived from local realism {\it plus auxiliary
assumptions. }Consequently the model violates the auxiliary assumptions,
although we will not discuss this point here in detail. At higher
efficiency, the model departs from the conventional quantum theory in that
it predicts a nonlinear response of the detectors (see eqs.$\left( \ref{dete}%
\right) $ and $\left( \ref{5a}\right) )$, or a high dark rate (see eq.$%
\left( \ref{dark}\right) ),$ or both. This is the feature that prevents the
violation of a genuine Bell's inequality (that is, one involving no
auxiliary assumptions in addition to local realism).

An obvious disproof of our LHV model would be achieved if a PDC experiment
violated a (genuine) Bell's inequality; this is what is usually called a
``loophole-free'' test. But, at its most optimistic, such a test, using
parametric down-converted photon pairs, lies in the remote future. In any
event, such a test would necessarily involve measurements of both single and
coincidence counts \cite{Santos}, and should avoid any background
subtraction. The latter condition derives from the fact that our model
predicts the existence of a fundamental dark rate in photon detectors and
there is no reason why it should satisfy a Bell's inequality if that rate is
subtracted.

\smallskip But simpler tests may be found for our model, since it has a
predictive power greater than that of quantum optics concerning the
behaviour of detectors. Nobody would claim that quantum mechanics had been
violated if it were discovered that there is a dark rate, or that the
response of the detector to the signal is non-linear, even though both of
these are in disagreement with the quantum theory of measurement. These
facts would be attributed to imperfect functioning of the detectors, and the
imperfections would be considered as technical problems of no relevance to
the testing of quantum theory. In sharp contrast, our model establishes
rather stringent fundamental constraints on the functioning of detectors,
precisely because the zeropoint field is taken as real. And the reality of
the ZPF is an unavoidable consequence of taking the Wigner function as the
probability distribution of hidden variables, which is the central idea of
our model.

The constraints posed by our model were already mentioned in section 7, and
they arise from the necessity of making compatible the conditions $\left( 
\ref{cond1}\right) $ and (\ref{cond2}), which imply 
\begin{equation}
\label{cond3}\overline{I}_{s}>>\sigma _{0}. 
\end{equation}
Hence, if we take into account (\ref{sigmacero}) and (\ref{icero}) we get 
\begin{equation}
\label{cond4}\overline{I}_{s}>>\frac{\hbar \overline{\omega }^{2}}{4cL\sqrt{%
\tau T}}. 
\end{equation}
This means that, in our detection model, there is a minimal effective
intensity of the signal which may be reliably detected, a constraint absent
in the quantum theory of detection. This is the constraint which may be put
to empirical test (remember that this constraint is required only if we
demand a low dark rate). Now we shall analyze the consequences of the
constraint.

As a consequence of the fact that the signal field has spatial coherence on
the surface of the lens, the intensity of the incident signal is amplified
by a factor 
$$
b^2\equiv \pi ^2R_l^4/\lambda ^2f^2, 
$$
$f$ being the focal distance. On the other hand, the zeropoint field is not
modified by the lens, which is evident because of the fact that energy
cannot be extracted from the vacuum.

$84\,\%$ ($91\,\%$) of the total intensity is concentrated within the first
(second) ring of the difraction pattern, with a radius $R=a\times (f\lambda
/2R_l)=a\lambda /A_r$, $A_r=2R_l/f$ being the relative aperture of the lens,
and $a=1.22$ ($2.23$) for the first (second ring) \cite{born}. Consequently,
the optimus radius of the detector is given by $R$. After that we may write
constraint $\left( \ref{cond4}\right) $ in terms of the intensity, I$_{IN}$,
arriving at the aperture of the detection system as follows 
$$
I_{IN}=b^{-2}\overline{I}_s>>\frac{\hbar \overline{\omega }^2\lambda ^2f^2}{%
4\pi ^2R_l^4cL\sqrt{\tau T}}. 
$$
Still, we may write the constraint in terms of the most direct empirical
quantity, namely the single counting rate, and we get 
\begin{equation}
\label{cond6}Rate=\frac{\eta \pi R_l^2I_{IN}}{\hbar \overline{\omega }}>>%
\frac{\eta \lambda f^2}{2R_l^2L\sqrt{\tau T}}, 
\end{equation}
where $\eta $ is the quantum efficiency of the detector and we have used the
equality $\overline{\omega }=2\pi c/\lambda .$ It may appear that the
detection rate could be as low as we want by just increasing the radius, R$%
_l $, of the lens, but this is not true because the condition $\left( \ref
{spc}\right) $ should be also fulfilled. If we combine this with $\left( \ref
{spc}\right) $ we obtain 
\begin{equation}
\label{cond7}Rate>>\frac{\eta f^2R_C^2}{2Ld^2\lambda \sqrt{\tau T}}. 
\end{equation}

The constraints $\left( \ref{cond6}\right) $ and $\left( \ref{cond7}\right)
, $ which put a lower bound to the single rate which may be used in reliable
experiments, is the most dramatic prediction of our model. The existence of
a lower bound to the single rate for the reliability of PDC experiments
cannot be derived from (conventional) quantum theory. If we put typical
parameters of the detector, that is $\eta \approx 0.1,$ $R_{l}$, $L$ and f
of order of fractions of a centimeter and a time window $T\approx 10\,$ns,
and use typical values of the wavelength, $\lambda $ $\approx 700\,$nm, and
bandwidth, $\Delta \lambda \approx 10$ nm (which gives a coherence time $%
\tau $ $\approx 1\,4$ ps) we get a minimal counting rate of the order of 10$%
^{5}-10^{6}$ counts per second. This figure is not far from the one
appearing in actual experiments. In any case, the model requires
improvements in order to be able to make more accurate predictions, a work
which is in progress.

\section{Appendix}

In order to demonstrate Eq. (\ref{noll}) we shall first substitute Eq. (\ref
{2a}) into Eq. (\ref{imed}). We shall express the electric field as the sum
of two terms (see Eq. (\ref{pdc})):

\begin{equation}
\label{a}E^{(+)}({\bf r},t;\alpha )=E_0^{(+)}({\bf r},t;\alpha )+E_s^{(+)}(%
{\bf r},t;\alpha ), 
\end{equation}
where $E_s^{(+)}({\bf r},t;\alpha )$ is the part of the field which is
superimposed to the zeropoint field $E_0^{(+)}({\bf r},t;\alpha ).$ The
calculation of the effective intensity corresponding to the filtered
zeropoint field has been computed in Eq. (\ref{icero}). Let us now focus on
the calculation of $\overline{I}_s$, which is given by the following
expression:

\begin{equation}
\label{b}
\begin{array}{c}
\overline{I}_s=\left( \frac 1{\pi R^2LT}\right) ^2c\varepsilon
_0\int_{V_{\det }}dV\int_{V_{\det }}dV^{\prime }\int_0^Tdt\int_0^Tdt^{\prime
}\langle E_s^{(+)}({\bf r},t;\alpha )E_s^{(-)}({\bf r}^{\prime },t^{\prime
};\alpha )\rangle \\ \times \left[ \sum_j\mbox{e}^{-i\frac{{\bf \omega }_j}%
c(z-z^{\prime })+i\omega _j(t-t^{\prime })}\right] . 
\end{array}
\end{equation}
The spatial coherence implies that the integrals over the surface of the
detector are equal to $\pi R^2$. Passing to the continuous in the summation
in $j$ Eq. (\ref{b}) transforms into

\begin{equation}
\label{c}
\begin{array}{c}
\overline{I}_s=\left( \frac 1{LT}\right) ^2c\varepsilon
_0\int_{-L/2}^{+L/2}dz\int_{-L/2}^{+L/2}dz^{\prime
}\int_0^Tdt\int_0^Tdt^{\prime }\langle E_s^{(+)}(z,t;\alpha
)E_s^{(-)}(z^{\prime },t^{\prime };\alpha )\rangle \\ \times \left[ \frac
T{2\pi }\int_{\omega _{\min }}^{\omega _{\max }}d\omega _j 
\mbox{e}^{-i\frac{{\bf \omega }_j}c(z-z^{\prime })+i\omega _j(t-t^{\prime
})}\right] \\ =\Delta \omega 
\frac{c\varepsilon _0}{L^2T^2}\frac T{2\pi
}\int_{-L/2}^{+L/2}dz\int_{-L/2}^{+L/2}dz^{\prime
}\int_0^Tdt\int_0^Tdt^{\prime }\langle E_s^{(+)}(z,t;\alpha
)E_s^{(-)}(z^{\prime },t^{\prime };\alpha )\rangle \\ \times \mbox{e}^{i%
\overline{\omega }(t-t^{\prime }-\frac{z-z^{\prime }}c)}\mbox{sinc}[\frac{%
\Delta \omega }2(t-t^{\prime }-\frac{z-z^{\prime }}c)]. 
\end{array}
\end{equation}
Now, we make the substitution $\mbox{sinc}[\frac{\Delta \omega }%
2(t-t^{\prime }-\frac{z-z^{\prime }}c)]\rightarrow \frac{2\pi }{\Delta
\omega }\delta (t-t^{\prime }-\frac{z-z^{\prime }}c),$ and perform one
integration on time. We have

\begin{equation}
\label{d}
\begin{array}{c}
\overline{I}_s=\frac{c\varepsilon _0}{L^2T}\int_{-L/2}^{+L/2}dz%
\int_{-L/2}^{+L/2}dz^{\prime }\int_0^Tdt\langle E_s^{(+)}(z,t;\alpha
)E_s^{(-)}(z^{\prime },t-\frac{z-z^{\prime }}c)\rangle \\ = 
\frac{c\varepsilon _0}{LT}\int_{-L/2}^{+L/2}dz\int_0^Tdt\langle
E_s^{(+)}(z,t;\alpha )E_s^{(-)}(z,t;\alpha )\rangle . \\ \Rightarrow
\varsigma \bar I_s=\frac{\eta A}{h\nu }\frac
1L\int_{-L/2}^{+L/2}dz\int_0^Tdt\ c\varepsilon _0\langle
E_s^{(+)}(z,t;\alpha )E_s^{(-)}(z,t;\alpha )\rangle . 
\end{array}
\end{equation}
This result coincides with the quantum one, Eq. (\ref{641}), when the
spatial coherence is taken into account in that equation.

\section{Acknowledgement}

We acknowledge financial support by DGICYT Project No. PB-95-0594 (Spain).

\end{document}